\documentclass[twocolumn]{aastex62}

\usepackage{savesym}
\usepackage{amsmath}
\savesymbol{iint}
\savesymbol{iiint}
\usepackage{txfonts}
\restoresymbol{TXF}{iint}
\restoresymbol{TXF}{iiint}

\def\ltsima{$\; \buildrel < \over \sim \;$}
\def\simlt{\lower.5ex\hbox{\ltsima}}
\def\gtsima{$\; \buildrel > \over \sim \;$}
\def\simgt{\lower.5ex\hbox{\gtsima}}

\def\ergs{{erg s$^{-1}$}}
\def\cm2{{cm$^{-2}$}}

\def\lum{{$L_{\rm X}$}}

\def\p1{{Paper I}}

\def\xmm{{\em XMM--Newton}}
\def\chandra{{\em Chandra}}

\def\swift{{\em Swift}}

\def\rosat{{\em ROSAT}}
\def\xmm{{\em XMM--Newton}}

\def\nh{{$N_{\rm H}$}}

\def\f14{{10$^{-14}$}}
\def\f13{{10$^{-13}$}}
\def\f12{{10$^{-12}$}}
\def\f11{{10$^{-11}$}}

\def\4u{{4U~1344$-$60}}

\def\feka{{Fe K$\alpha$}}

\def\lbol{{$L_{\rm Bol}$}}

\def\msun{{$M_{\rm \odot}$}}

\def\nus{{\em NuSTAR}}

\def\mbh{$M_{BH}$}
\def\ledd{$L_{\rm Edd}$}
\def\edd{$\lambda_{\rm Edd}$}

\def\f2{$f_{\rm 2}$}
\def\ecut{$E_{\rm cut}$}

\graphicspath{{./}{figures/}}

\received{March 26, 2019}
\revised{April 4, 2019}
\accepted{April 4, 2019}
\submitjournal{ApJL}

\shorttitle{Coronal properties of QSOs}
\shortauthors{Lanzuisi et al.}

\begin{document}

\title{NuSTAR MEASUREMENT OF CORONAL TEMPERATURE IN TWO LUMINOUS, HIGH REDSHIFT QSOs}

\correspondingauthor{Giorgio Lanzuisi}
\email{giorgio.lanzuisi@inaf.it}

\author[0000-0001-9094-0984]{G. Lanzuisi}
\affil{INAF- Osservatorio di Astrofisica e Scienza dello Spazio di Bologna, via Gobetti 93/3, 40129 Bologna, Italy}
\affil{Dipartimento di Fisica e Astronomia dell'Universit\'a degli Studi di Bologna, via P. Gobetti 93/2, 40129 Bologna, Italy}

\author{R.~Gilli}
\affiliation{INAF- Osservatorio di Astrofisica e Scienza dello Spazio di Bologna, via Gobetti 93/3, 40129 Bologna, Italy}

\author{M.~Cappi}
\affiliation{INAF- Osservatorio di Astrofisica e Scienza dello Spazio di Bologna, via Gobetti 93/3, 40129 Bologna, Italy}

\author{M.~Dadina}
\affiliation{INAF- Osservatorio di Astrofisica e Scienza dello Spazio di Bologna, via Gobetti 93/3, 40129 Bologna, Italy}

\author{S.~Bianchi}
\affiliation{Dipartimento di Matematica e Fisica, Universit\'a degli Studi Roma Tre, via della Vasca Navale 84, I-00146 Roma, Italy}

\author{M.~Brusa}
\affil{Dipartimento di Fisica e Astronomia dell'Universit\'a degli Studi di Bologna, via P. Gobetti 93/2, 40129 Bologna, Italy}
\affiliation{INAF- Osservatorio di Astrofisica e Scienza dello Spazio di Bologna, via Gobetti 93/3, 40129 Bologna, Italy}

\author{G.~Chartas}
\affil{Department of Physics and Astronomy of the College of Charleston, Charleston, SC 29424, USA}

\author{F.~Civano}
\affil{Harvard-Smithsonian centre for Astrophysics, 60 Garden Street, Cambridge, MA 02138, USA}

\author{A.~Comastri}
\affiliation{INAF- Osservatorio di Astrofisica e Scienza dello Spazio di Bologna, via Gobetti 93/3, 40129 Bologna, Italy}

\author{A.~Marinucci}
\affiliation{Dipartimento di Matematica e Fisica, Universit\'a degli Studi Roma Tre, via della Vasca Navale 84, I-00146 Roma, Italy}

\author{R.~Middei}
\affiliation{Dipartimento di Matematica e Fisica, Universit\'a degli Studi Roma Tre, via della Vasca Navale 84, I-00146 Roma, Italy}

\author{E.~Piconcelli}
\affiliation{INAF - Observatorio Astronomico di Roma, via Frascati 33, I-00040, Monte Porzio Catone, Roma, Italy}

\author{C.~Vignali}
\affil{Dipartimento di Fisica e Astronomia dell'Universit\'a degli Studi di Bologna, via P. Gobetti 93/2, 40129 Bologna, Italy}
\affiliation{INAF- Osservatorio di Astrofisica e Scienza dello Spazio di Bologna, via Gobetti 93/3, 40129 Bologna, Italy}

\author{W. N. Brandt}
\affil{Department of Astronomy and Astrophysics, 525 Davey Lab, The
     Pennsylvania State University, University Park, PA 16802, USA}
\affil{Institute for Gravitation and the Cosmos, The Pennsylvania State
     University, University Park, PA 16802, USA}
\affil{Department of Physics, 104 Davey Laboratory, The Pennsylvania State
     University, University Park, PA 16802, USA}

\author{F. Tombesi}
\affil{Department of Physics, University of Rome ‘Tor Vergata’, Via della Ricerca Scientifica 1, I-00133 Rome, Italy}
\affil{Department of Astronomy, University of Maryland, College Park, MD 20742, USA}
\affil{NASA/Goddard Space Flight Center, Code 662, Greenbelt, MD 20771, USA}
\affiliation{INAF - Observatorio Astronomico di Roma, via Frascati 33, I-00040, Monte Porzio Catone, Roma, Italy}

\author{M. Gaspari}\thanks{\textit{Spitzer} Fellow}
\affil{Department of Astrophysical Sciences, Princeton University, 4 Ivy Lane, Princeton,
NJ 08544-1001, USA}

\begin{abstract}
X-ray emission from AGN is believed to be produced via Comptonization of optical/UV seed photons emitted by the accretion disk, up-scattered by hot electrons in a corona surrounding the black hole. A critical compactness vs. temperature threshold is predicted above which any increase in the source luminosity, for a fixed size, would then generate positron-electron pairs rather than continue heating the coronal plasma.
Current observations seem to confirm that all AGN populate the region below this critical line.
These models, however, have never been probed by observations in the high-luminosity regime, where the critical line is expected to reach low temperatures. To fill this observational gap, we selected two luminous (log(\lbol)$>47.5$ erg/s) quasars,
2MASSJ1614346+470420 (z=1.86) and B1422+231 (z=3.62), and obtained \xmm\ and \nus\ deep observations for them. We performed detailed spectral analysis of their quasi-simultaneous soft and hard X-ray data, in order to constrain the parameters of their coronae.
Using a phenomenological cut-off power-law model, with the inclusion of a reflection component, we derived rest-frame values of the high energy cut-off of \ecut$=106^{+102}_{-37}$ keV and \ecut$=66^{+17}_{-12}$ keV, respectively. Comptonization models consistently give as best-fit parameters electron temperatures of $\sim45$ keV and $\sim28$ keV, respectively, and
optically thick coronae ($\tau>1$). 
These low coronal temperatures fall in the limited region allowed at these luminosities to avoid runaway pair production.
\end{abstract}


\section{Introduction} \label{sec:intro}

The primary X-ray emission in active galactic nuclei (AGN) is believed to be produced via Comptonization: optical/UV thermal photons emitted by the accretion disk are 
up-scattered by electrons in a hot ($\sim10^{8-9}$ K) corona surrounding the super-massive black hole (SMBH) (Haardt \& Maraschi 1993).
The resulting spectrum can be described as a cut-off power-law with a photon index and a high-energy cut-off (at energies around a few hundred keV) that depends on the electron temperature ($kT_e$) and the optical depth ($\tau$) of the corona.

The characterization of the X-ray emission from AGN is therefore the best tool available to investigate the physical properties of the innermost regions around accreting SMBHs and to measure coronal properties such as temperature, optical depth and geometry. Recent X-ray reverberation studies (De Marco et al. 2013, Reis \& Miller 2013, Cackett et al. 2014, Kara et al. 2016) suggest that the size $R_c$ of the hot corona producing the X-rays is in the range $\sim3-20$ $R_g$, where $R_g=GM/c^2$ is the gravitational radius for a black hole of mass M. Such a range has also been independently confirmed via microlensing studies (e.g. Pooley et al. 2007, MacLeod et al. 2015, Chartas et al. 2016).

Since interactions between high-energy photons in compact systems produce electron-positron pairs, it was soon realized that pair production in AGN coronae may act as an effective thermostat (e.g., Svensson 1984, Stern et al. 1995). In fact, any rise in electron temperature $T_e$ or compactness $\ell \propto L/R$ (Cavaliere \& Morrison 1980)\footnote{The dimensionless compactness $\ell$ is defined as the luminosity L over the size R of the emitting region, $\ell=L \sigma_T/R m_e c^3$} above a critical curve in the temperature-compactness plane, the so called ``pair line'', would result in a runaway pair production, causing the temperature to drop.

Early studies were based on high-energy X-ray missions such as {\em CGRO}, {\em BeppoSAX}, {\em Integral}, \swift-BAT and Suzaku, mounting non-imaging instruments whose ability to obtain high S/N spectra and measure spectral cut-offs were limited to bright nearby sources ($F_x > 10^{-11}$ ergs s$^{-1}$ cm$^{-2}$). Yet, they showed that local Seyfert galaxies exhibit coronae with a broad range of temperatures (\ecut=50-500 keV, Perola et al. 2002, Dadina et al. 2008, Vasudevan et al. 2013, Malizia et al. 2014) in the luminosity range \lum$=10^{42}-10^{44}$ erg/s.

%%%%%%%%%%%%%%%%%%%%%%%%%%%%%%%%%%%%%%%%%%%%%%%%%%%%%%%%%%%%%%%%%%%%%%%
\begin{table*}[t]
\begin{center}
\caption{Target properties} 
\renewcommand{\arraystretch}{1.2}
\begin{tabular*}{\textwidth}{@{\extracolsep{\fill} }l c c c c c c c}
\noalign{\smallskip} \hline \hline \noalign{\smallskip}
Target                    &$ z_{spec}$&   $\mu$ & \lbol          & log$M_{BH}$   & Edd.   & \xmm\    & \nus \\
                          &           &         &  (erg/s)       & \msun         &        & (ks)     &   (ks) \\
    (1)  & (2) & (3) & (4) & (5) & (6)  & (7) & (8) \\                        
\hline \noalign {\smallskip}
B1422               &  3.62    &    20$^a$  &    47.15$^a$   &  9.7$^a$      &  0.21  & 25     &   85  \\      
2MASSJ16        & 1.86     &      --    &    47.79$^b$   & 9.8$^b$       &  0.71  & 69     &   140   \\
\hline \noalign {\smallskip}
\end{tabular*}
\end{center}
\vspace{-0.3cm}
{\bf Notes.} (1) Target Name, (2) Redshift,  (3) Lensing factor (4) Log. of the Bolometric luminosity, (5) Log. of the SMBH mass in \msun, (6) Eddington ratio, defined as \lbol/\ledd, (7) \xmm-pn cleaned exposure time, (8) \nus\ cleaned exposure time. $^a$ Assef et al. (2011), \lbol\ derived from the de-lensed $L_{1450\AA}$ (20\% conservative error estimate), $M_{BH}$ from H$_{\beta}$. $^b$ Shen et al. (2011), $M_{BH}$ from Mg\,{\sc ii}.
\end{table*}
%%%%%%%%%%%%%%%%%%%%%%%%%%%%%%%%%%%%%%%%%%%%%%%%%%%%%%%%%%%%%%%%%%%%%%%

\nus\ (Harrison et al. 2013) is now providing a major advance in the understanding of AGN coronae. Its bandwidth is limited to  $\sim80$ keV, but its sensitivity is orders of magnitude better than previous missions, allowing accurate measurements of the cut-off in local AGN.
Fabian et al. (2015, F15 hereafter) compiled a list of all high-energy cut-offs measured by \nus\, and investigated the $\ell$ vs. $T_e$ relation in detail. The temperature was derived assuming \ecut$/k T_e\sim2$ as found by Comptonization models for optically thin coronae (e.g. Petrucci et al. 2001) and $R_c$ was generally assumed to be $10 R_g$. Most sources were found to be below the $\ell$ vs. $T_e$ critical line defining the region forbidden by the onset of pair production.

In general it is difficult to tightly constrain cut-off values exceeding the observed band pass
(Parker et al. 2015) except for sources with very high photon statistics (see e.g. Matt et al. 2015, Garc{\'\i}a et al. 2015).
\nus\ measurements, all performed so far in nearby ($z < 0.06$) low luminosity log\lum$< 45$ erg/s AGN, are in fact limited by the \nus\ band-pass (\ecut$<200$ keV).

Therefore the high-\lum, high-\ecut\ regime has never actually been probed. 
However, thanks to its greater sensitivity, \nus\ is now capable of testing runaway
pair-production models by measuring coronal properties in high redshift, high luminosity QSOs where larger \ecut\ values can be constrained thanks to the cosmological redshifting of the cut-off downward in observed-frame energy. 

We present here the first firm measurement of \ecut\ in two high-redshift ($z\simgt2$), high-luminosity (\lbol$>47$ \ergs) QSOs.  The paper is organized as follows:
Sec.~2 describes the target selection, Sec.~3 the data reduction.  In Sec.~4 we report the different spectral models adopted and in Sec.~5 we summarize our results.
We adopt the cosmological parameters $H_0 = 70$ km s$^{-1}$ Mpc$^{-1}$, $\Omega_{\Lambda} = 0.73$ 
and $\Omega_m = 0.27$. Errors  are given at 90\% confidence level.

\section{Target selection and observations} \label{sec:selection}

In order to select luminous QSOs, bright enough in the X-ray band to allow for a good characterization of the high energy cut-off, we considered all known QSOs with 
$F_{0.5-10 keV}~>~5\times~10^{-13}$ ergs s$^{-1}$ cm$^{-2}$ at spectroscopic redshift $z_{spec} > 1.5$, both lensed and non-lensed sources. 

%%%%%%%%%%%%%%%%%%%%%%%%%%%%%%%%%%%%%%%%%%%%%%%%%%%%%%%%%%%%%%%%%%%%
\begin{figure*}[t]
\begin{center}

\includegraphics[width=7.7cm]{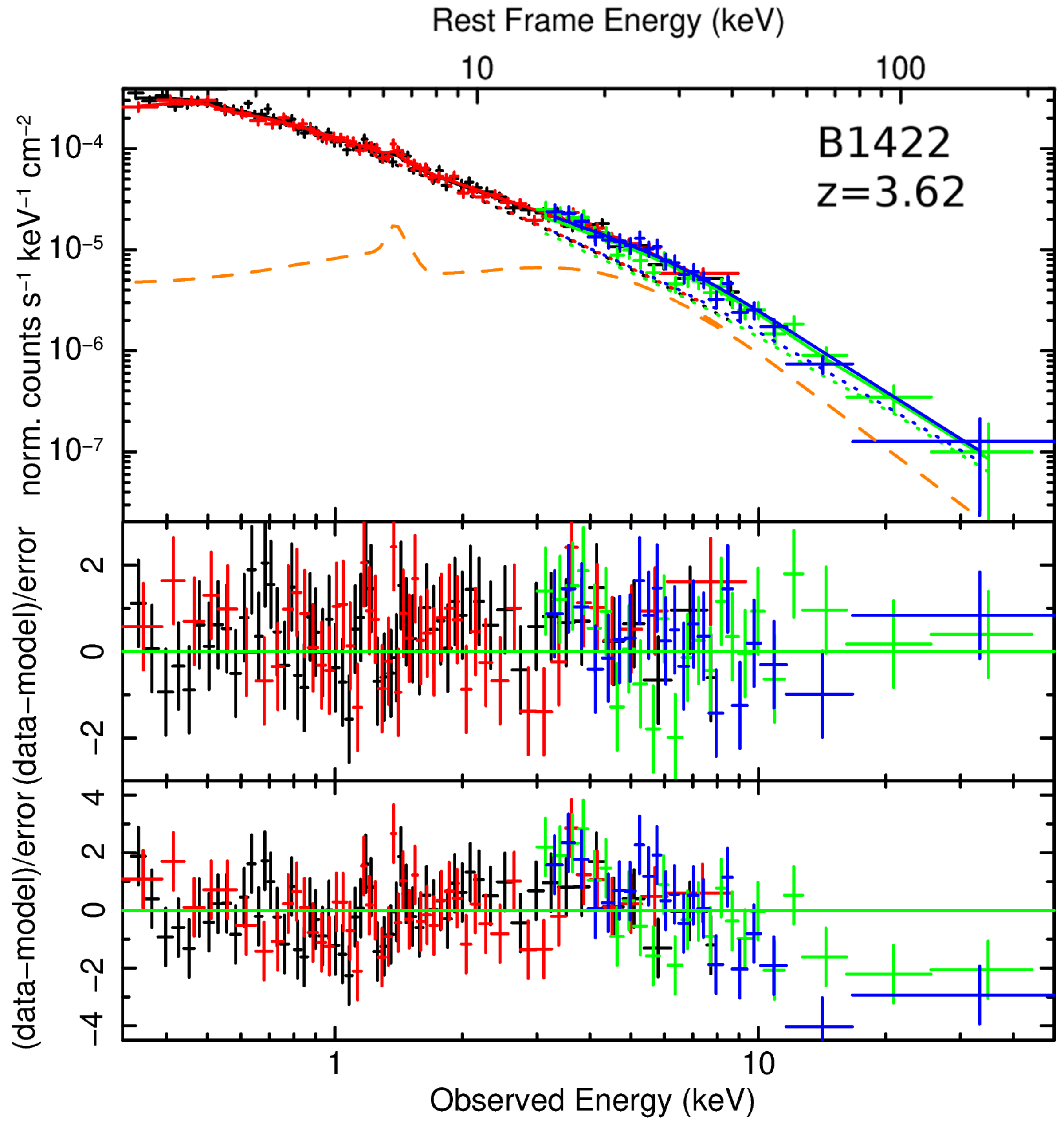}\hspace{0.7cm}\includegraphics[width=7.7cm]{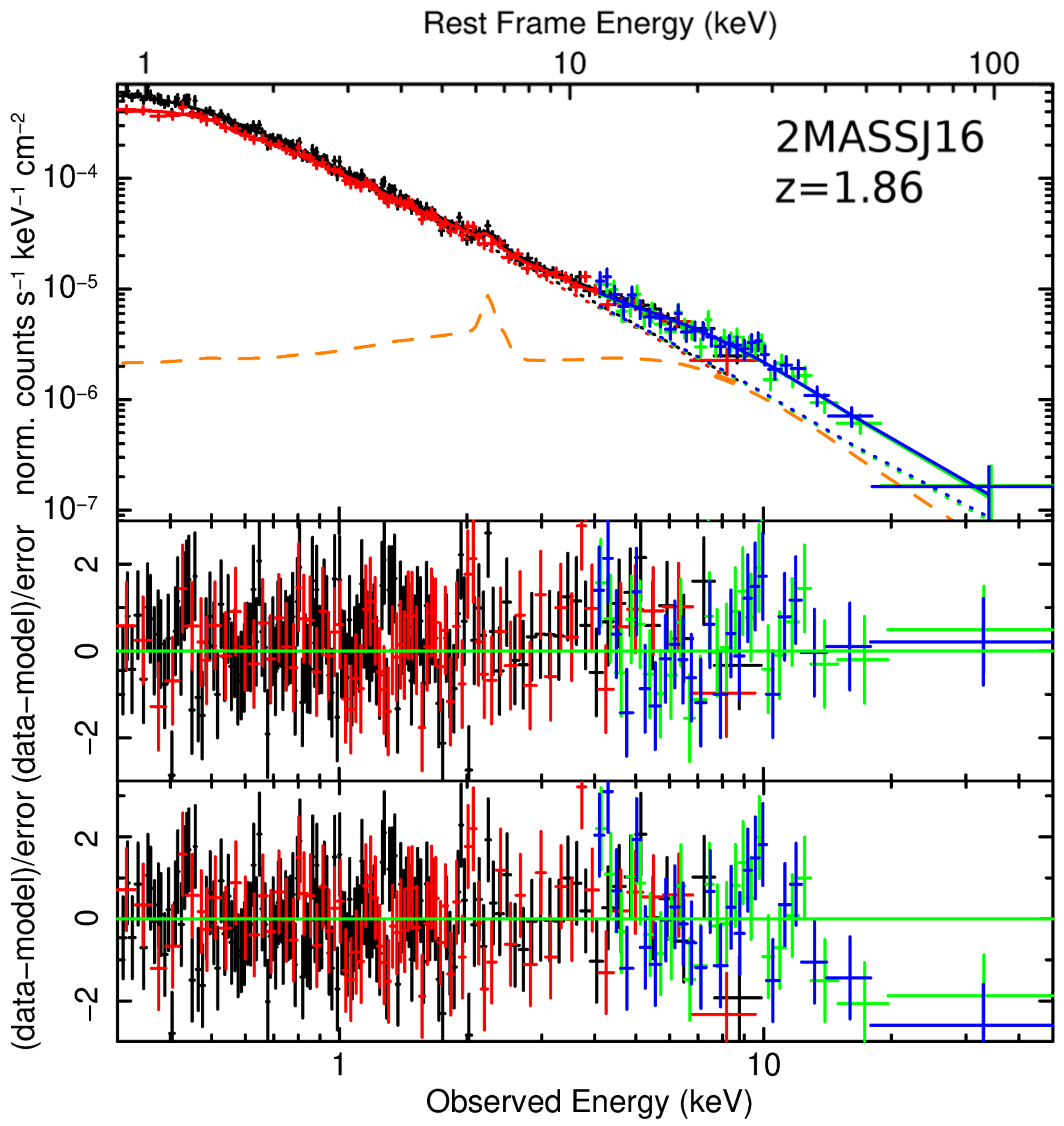}\vspace{0.2cm}
\includegraphics[width=7.8cm]{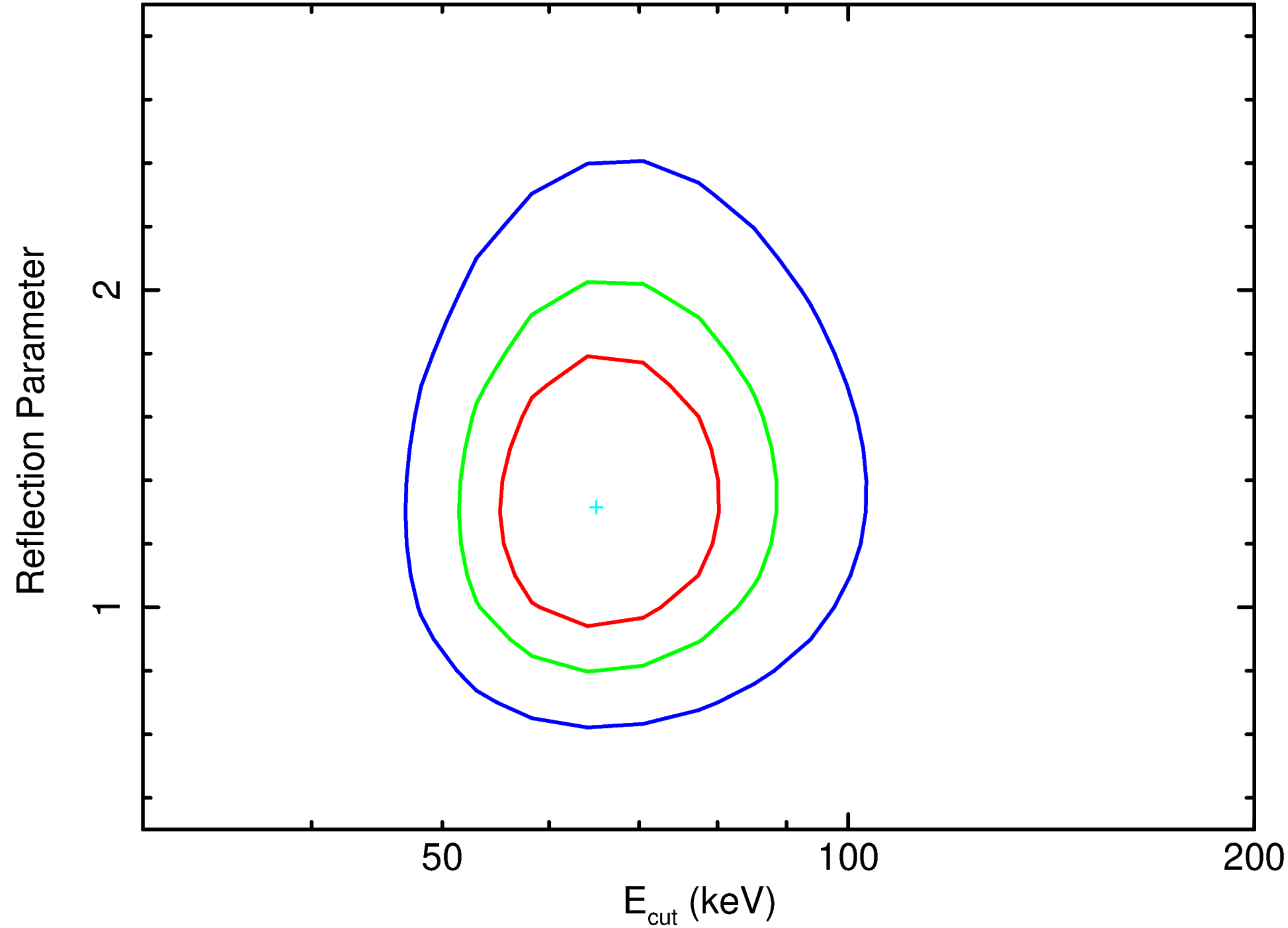}\hspace{0.5cm}\includegraphics[width=7.7cm]{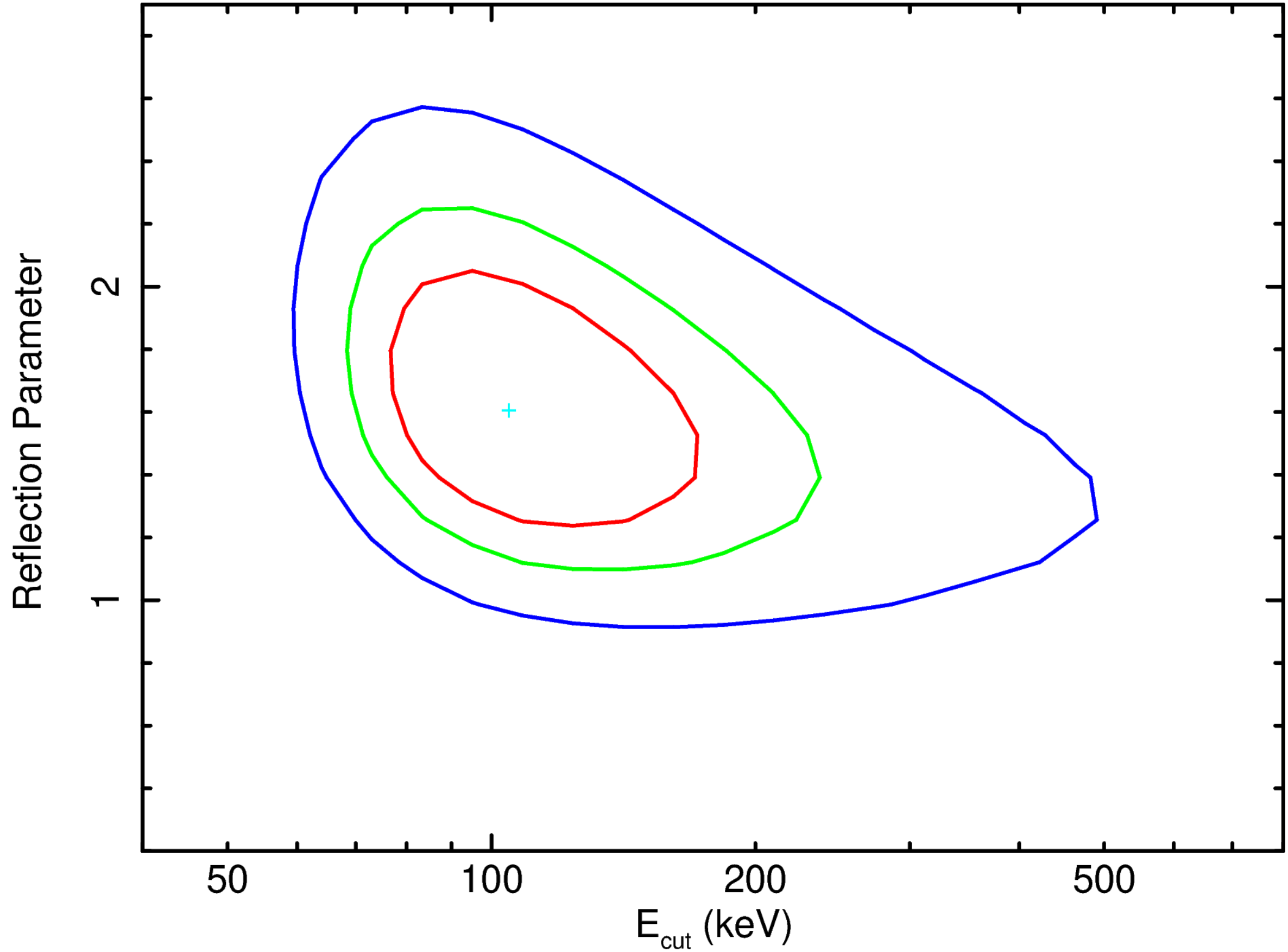}\vspace{0.2cm}
\includegraphics[width=7.9cm]{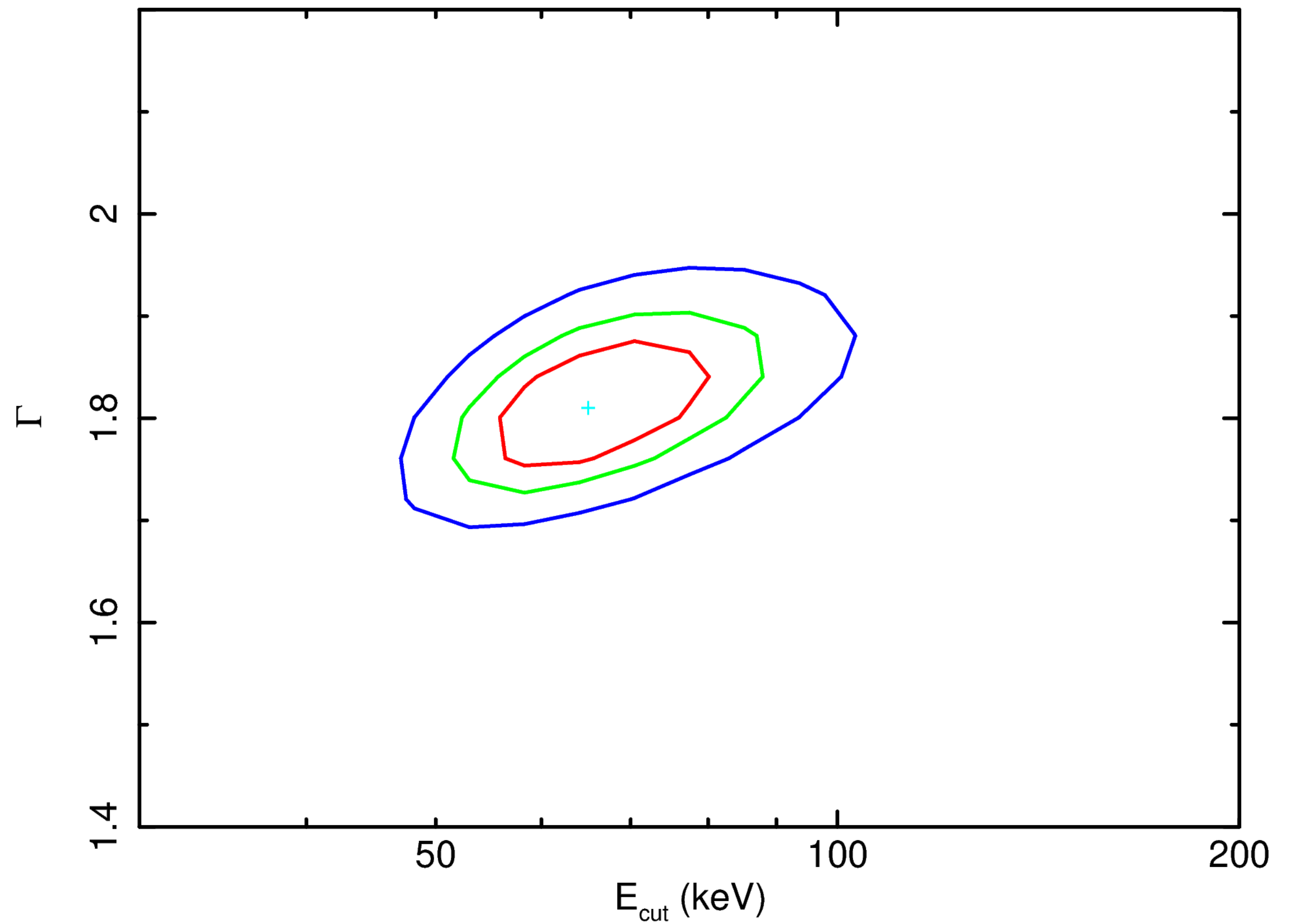}\hspace{0.45cm}\includegraphics[width=7.7cm]{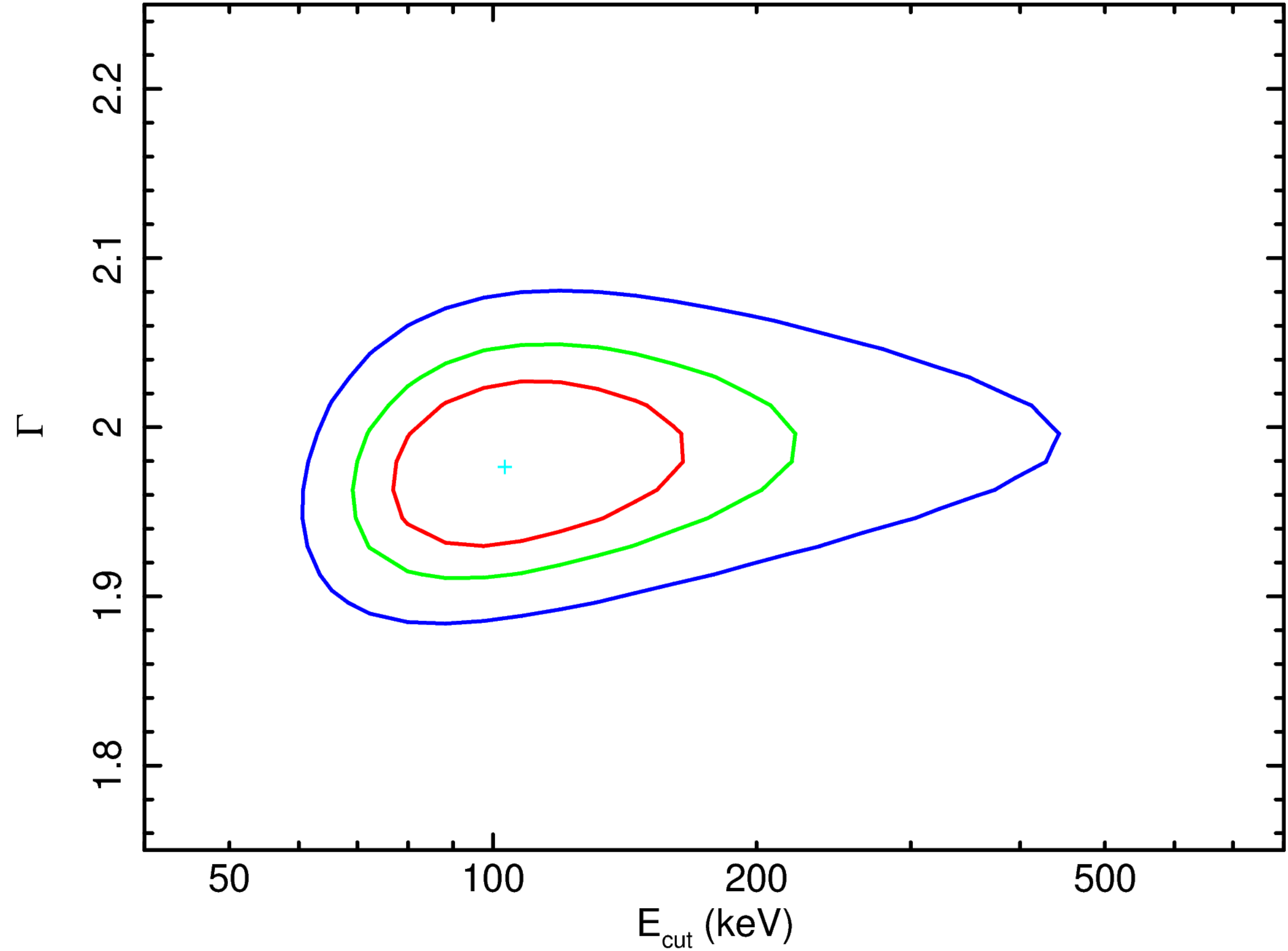}
\caption{{\it Top:}
Normalized spectra and residuals from the fit of B1422 (left) and 2MASSJ16 (right) with the phenomenological model. The lower panels show residuals from the fit with no high-energy cut-off. Black, red, green and blue points show \xmm\ pn, MOS1+2, \nus\  FPMA and FPMB data, respectively. Data binned for plotting purposes. The orange dashed lines show the reflection component.
{\it Center:}
Confidence contours, at 68, 90 and 99\% c.l., of the reflection parameter R vs. \ecut.
{\it Bottom:}
Same for the photon index $\Gamma$ vs. \ecut.}
\label{fig:spectra}
\end{center}
\end{figure*}
%%%%%%%%%%%%%%%%%%%%%%%%%%%%%%%%%%%%%%%%%%%%%%%%%%%%%%%%%%%%%%%%%%%%%%%

Lensed AGN were selected from the CASTLES catalog\footnote{See \url{https://www.cfa.harvard.edu/castles}}. The brightest lensed source in this catalog is B1422+231 (B1422 hereafter). The lensing factor is estimated to be $\sim20$ (Assef et al. 2011).

As for non lensed sources, we searched for the brigthest QSOs by cross-correlating 
the X-ray point-source catalogs from \chandra, \xmm\ and \rosat\ (CSC v2, 3XMM-DR6 and RASS-BSC), with the 12th SDSS-III data release. The brightest one is 2MASSJ1614346+470420 (2MASSJ16 hereafter).
These two sources (B1422 and 2MASSJ16) were observed quasi-simultaneously with \xmm\ and \nus\ in 2017 as part of a \nus\ Cycle 3 program (PI Lanzuisi)\footnote{B1422 has also been observed with \chandra\ several times, for a total of 125ks, the most recent one being in 2012. In order to avoid long-term variability issues (see e.g. Lanzuisi et al. 2016) we focus our analysis on the coeval \xmm\ and \nus\ data.}. 
Their properties are summarized in Table 1. 

We note that B1422 is classified as moderately radio loud ($R=90$, Dadina et al. 2016) with a steep radio continuum ($\alpha_r=0.9$, Orienti et al. 2007), indicating that the source is highly inclined in the plane of the sky. In
this case, the radio emission should be dominated by the lobes, and not by the jet, and the X-ray spectrum is not strongly contaminated by the jet component (see discussion in Dadina et al. 2016).

\section{Data reduction} \label{sec:data}

B1422 was observed by \xmm\ on 2017-12-29 for 38 ks and by \nus\ on 2017-12-30 for 101 ks. 2MASSJ16 was observed by \xmm\ on 2017-08-05 for 98 ks, and followed up by \nus\ on 2017-08-28 and 2017-10-09 for 106 and 48ks, respectively.

For both sources, \xmm\ EPIC data were reduced using the standard software
\texttt{SAS} v.16.1\footnote{\url{https://www.cosmos.esa.int/web/xmm-newton/sas}}.
A filter for periods of high background rate was adopted using a threshold 
of 0.5 and 0.2 counts per seconds in the 10-12 keV band, for pn and MOS, respectively. 
We selected only events corresponding to single and double pixel events (pattern 0-4 and 0-12 for pn and MOS respectively). The source spectra were extracted from circular regions of 40'' radius, corresponding to $\sim90\%$ encircled energy fraction. The background spectra were extracted from an area $\sim10$ times larger than the source region, surrounding the QSO. The final exposure times were 25ks pn (35ks MOS) for B1422 and 69ks pn (84ks MOS) for 2MASSJ16.

The \nus\ data were processed using the \nus\ Data Analysis Software package (NuSTARDAS) v.1.8.0 within \texttt{Heasoft} v. 6.20  tools\footnote{\url{https://heasarc.nasa.gov/lheasoft/}}.
Calibrated and cleaned event files were produced using the calibration files in the \nus\ CALDB (version 20170727) and standard filtering criteria with the \texttt{Nupipeline} task. We checked for high background period using the \texttt{nustar\_filter\_lightcurve} IDL script\footnote{\url{https://github.com/NuSTAR/nustar-idl}}. The two \nus\ observations for 2MASSJ16 were taken at 40 days distance, and we verified that no significant variability was detected before merging the two data-sets with standard \texttt{Heasoft} tools. The final, cleaned \nus\ exposure times are 85ks for B1422 and 140ks for 2MASSJ16.

In order to reduce the background and increase the spectral signal to noise at high energies, we tested different extraction regions, and finally adopted a region of 40'' radius, corresponding to $\sim60\%$ of the encircled energy fraction for the \nus\ PSF 
(An et al. 2014). This allowed us to detect the source at $>3\sigma$ in the $20-50$ keV band in both sources, and better sample the high-energy range.

\section{Spectral modeling} \label{subsec:model}

The final spectra have $1.2\times10^4$ and $3.0\times10^4$ total, 0.3-10~keV \xmm\ counts and 1800 and 2300 total, 3-50 keV \nus\ counts for B1422 an 2MASSJ16, respectively.
The spectral modeling is performed with the package \texttt{Xspec} v. 12.9.1, using the the C-stat statistic (Cash 1979), and binning the spectra to 5 counts per bin, since \nus\ spectra are in the low-counts regime. 
Given the quality of the available data we choose not to include in our analysis complex models such as relativistic reflection.
All the models described below are modified by a Galactic column density of \nh$= 3.2\times10^{20}$ \cm2\ for B1422 and $0.9\times10^{20}$ \cm2 for 2MASSJ16, respectively (Kalberla et al. 2005).

\subsection{Phenomenological model}

As a first step we fitted the \xmm\ and \nus\ spectra for both sources with a phenomenological model:  a power-law with an exponential high-energy cut-off and Compton reflection from cold material in a slab geometry, including emission lines (model \texttt{Pexmon} in \texttt{Xspec}, Nandra et al. 2007). The intensity of the reflection is parametrized with R, defined as the solid angle covered by the cold, reflecting material, as visible from the Comptonizing source, in units of $2\pi$.

Given the possible degeneracy between spectral slope, intensity of the reflection and high-energy cut-off (Perola et al. 2002), all these component must be fitted simultaneously. 
The free parameters of the model are, therefore, the power-law photon index $\Gamma$; the high-energy cut-off \ecut; the reflection parameter R and the continuum normalization.
In neither source intrinsic cold absorption in addition to the Galactic value is required.

The model is the same for all the four data sets (pn, MOS1+MOS2\footnote{MOS1 and 2 spectra merged, and response matrices averaged with standard HEASARC \texttt{ftools}, \url{http://heasarc.gsfc.nasa.gov/ftools/}.} and \nus\ FPMA and FPMB) and a flux cross-calibration $C_i$ is applied between the different instruments with values always smaller than $1.15$ (the spectral slopes $\Gamma$ obtained fitting each instrument separately are consistent within errors).

We fixed all element abundances to solar values and fixed the inclination angle to $i=60^\circ$. 
We tested that adopting a different inclination angle has a limited impact on the resulting \ecut\ (few \% difference for $i=30^\circ$ and $i=80^\circ$), while it has a strong impact on the reflection parameter: R is a factor $\sim2$ lower (higher) for $i=30^\circ$ ($i=80^\circ$).

As can be seen from Fig.~1 (top left and right), the phenomenological model is able to fully reproduce the broa-band spectrum of both QSOs, and the \nus\ data are crucial to constrain the high-energy cut-off of the continuum. The best-fit continuum parameters are summarized in Tab.~2. 

%%%%%%%%%%%%%%%%%%%%%%%%%%%%%%%%%%%%%%%%%%%%%%%%%%%%%%%%%%%%%%%%%%%%%%%
\begin{table*}[t!]
\begin{center}
\caption{Best-fit Parameter Values for the \texttt{pexmon} Model}	\renewcommand{\arraystretch}{1.2}
\begin{tabular*}{\textwidth}{@{\extracolsep{\fill} }l c c c c c c c c}
\noalign{\smallskip} \hline \hline \noalign{\smallskip}
Target                    & $\Gamma$ &  \ecut\   & R    &   $F_{0.5-10}$       &   log$L_{2-10}$ & C$_{MOS}$ & C$_{FPMA}$/C$_{FPMB}$  & $Cstat/d.o.f.$  \\
                          &          &   keV     &      &    ($10^{-13}$ cgs)  &    (erg/s)   &       \\
(1)  & (2) & (3) & (4) & (5) & (6)  & (7) & (8) & (9) \\                        
\hline \noalign {\smallskip}
B1422                 &  $1.81_{-0.06}^{+0.07}$ & $66_{-12}^{+17}$  & $1.3_{-0.4}^{+0.5}$ &  $9.5$  &   45.30    &  1.03 & 1.02/1.11 & 1637/1668      \\      
2MASSJ16      &  $1.98_{-0.05}^{+0.11}$ & $106_{-37}^{+102}$ & $1.6_{-0.5}^{+0.7}$ &  $6.7$ &   45.97  & 0.94 &  1.09/1.12 & 1705/1735   \\
\hline \noalign {\smallskip}
\end{tabular*}
\end{center}
\vspace{-0.3cm}
{\bf Notes.} (1) Target Name, (2) photon index, (3) high energy cut-off in keV, (4) reflection parameter, (5) Observed \xmm\ 0.5-10 keV flux in erg s$^{-1}$ cm$^{-2}$, (6) Log. of the intrinsic (de-lensed) 2-10 keV luminosity in erg/s, (7) \xmm\ MOS cross calibration with respect to pn, (8) \nus\ FPMA/FPMB cross-calibration with respect to pn, (9) $Cstat/d.o.f.$ of the best-fit.
\end{table*}
%%%%%%%%%%%%%%%%%%%%%%%%%%%%%%%%%%%%%%%%%%%%%%%%%%%%%%%%%%%%%%%%%%%%%%%

Performing the fit with the \texttt{Pexrav} model (Magdziarz \&  Zdziarski 1995) plus the \feka\ emission line, we get consistent best-fit results for the continuum parameters. In addition, in B1422 the line is detected at 90\% c.l., with parameters $E_{line}=6.50_{-0.14}^{+0.18}$ keV and $EW=80_{-40}^{+60}$ eV rest frame (consistent with the results obtained in Dadina et al. 2016 from a deeper \xmm\ observation), while in 2MASSJ16 it is marginally detected, with $EW<125$ eV.

Confidence contours for $\Gamma$ and R  vs. \ecut\ are shown in Fig.~1 (lower panels). All three continuum parameters are well constrained. The photon index is typical of QSOs (Piconcelli et al. 2005), while the reflection parameter is higher than the one measured by \nus\ in luminous QSOs (Del Moro et al. 2017, Zappacosta et al. 2018): at \lum$>10^{45}$ erg/s the typical reflection parameter is $R<1$.

Finally, the \ecut\ values are lower than the average observed for low luminosity AGN: \ecut$=66_{-12}^{+17}$ keV and \ecut$=106_{-37}^{+102}$ keV for B1422 and 2MASSJ16, respectively, implying coronal temperatures of $\sim33$ keV and $\sim53$ keV, assuming \ecut$/kT_e\sim2$ for optically thin coronae (Petrucci et al. 2001). For B1422, our best-fit \ecut\ is consistent, within $1\sigma$, with the results from Dadina et al. (2016), but the uncertainties derived here are 50\% smaller. 

These \ecut\ values place the two QSOs in the narrow region allowed by the runaway pair-production models: in Fig.~2 (left) we translated the $\ell$ vs. $T_e$ plane from F15 in terms of directly observable quantities, i.e. \lum\ vs. \ecut. The critical \lum\ vs. \ecut\ lines are obtained from the theoretical critical $\ell$ vs. $T_e$ lines of Stern et al. (1995) for the two assumed coronal geometries, i.e. a slab or a hemisphere above the accretion disk. In the conversion we assumed 
 $R_c = 10 R_g$ and \mbh$ = 10^8$ \msun\ (the mean mass of the F15 sample, thick lines) and \mbh$ = 10^9$ \msun\ (as appropriate for luminous and massive QSOs, dashed lines).
 
We updated the \nus\ compilation of F15 by adding six more AGN with measured \ecut\ from recent literature (Tortosa et al. 2017, Kammoun et al. 2017, Buisson et al. 2018, Younes et al. 2019). We also updated, for nine sources, the values 
of \ecut\ obtained in Malizia et al. (2014) with Integral, with the ones derived with \nus\ in Molina et al. (2019). The new \ecut\ values are in very good agreement with the previous ones, with narrower error bars.
Finally, we added the results from Ricci et al. (2018) from a large sample of local AGN observed with \swift--BAT, binned in compactness. Again we used $R_c = 10 R_g$ and \mbh$ = 10^8$ \msun\ to convert from physical to observable quantities.

Most sources in Fig.~2 (left) lie where the runaway pair-production model predicts. Some are close to the critical lines (NGC 5506 being the highest \nus-constrained \ecut), but always below the theoretical limits for pair balance, suggesting that pair-production acts to regulate the coronal temperature. 
The two QSOs analyzed here strikingly fall in the restricted region of high luminosity-low temperature allowed by the model.

%We stress that this result is not affected by the observational bias related to the difficulty of measuring higher values of \ecut: we have tested with simulations that, with the data quality available and at the redshift of the two QSOs, we would have been able to firmly measure \ecut\ as high as $\sim500\pm200$ keV rest frame.

Finally, we note that the observed \ecut\ should be corrected for the effect of gravitational redshift. This can be parametrized with the $g$-factor $g=$ \ecut$^i$/\ecut$^o$. Tamborra et al. (2018a) computed how $g$ depends on the radius/height of the corona, its emissivity profile, the BH spin and the system inclination. For $R_c=10$ and most combinations of these parameters, $g$ falls between 1 and 1.18, being at most 1.38 for a maximally spinning BH seen at $30^{\circ}$ with steep radial emissivity profile ($r^{-3}$). The redshifting of relativistic reflection may also contribute to the shift of \ecut$^o$ to a lesser extent.

\subsection{Comptonization model}

We tested also the physical, Comptonization model \texttt{NthComp} in Xspec (Zycki, Done \& Smith 1999). In order to properly take into account also the non negligible reflection component derived with the phenomenological model, we used the \texttt{xillverCp} reflection model (Garcia et al. 2014) that uses \texttt{NthComp} as primary emission and self-consistently compute reflection from a non-relativistic medium. The free parameters are the electron temperature $kT_e$, the photon index $\Gamma$, the ionization parameter of the reflector, and the reflection fraction. The optical depth $\tau$ can be derived indirectly from $kT_e$ and $\Gamma$ (Zdziarski et al. 1996). 

The best-fit model, with comparable $Cstat/d.o.f.$ of the phenomenological model, has very low electron temperatures of $kT_e=22_{-2}^{+7}$ keV for B1422 and $kT_e=42_{-11}^{+10}$ keV for 2MASSJ16, and optical depths significantly above unity, $\tau=3.6_{-0.4}^{+0.6}$ and $\tau=2.0_{-0.2}^{+0.3}$, respectively. The ionization parameter of the reflector is consistent with 0, and the reflection parameter is similar to the one derived with the phenomenological model ($R=1.4-1.8$).

We further tested these results using the Monte Carlo Comptonization model MoCA (Tamborra et al. 2018b, Marinucci et al. 2018). The $kT_e$ and $\tau$ obtained with MoCA, and derived from the photon index and \ecut\ results from the phenomenological model, are $kT_e=26_{-3}^{+8}$ keV and $\tau=3.2_{-0.8}^{+0.6}$ for B1422, and 
$kT_e=44_{-11}^{+26}$ keV and $\tau=1.6_{-0.7}^{+0.8}$ for 2MASSJ16 in the slab geometry. 
The spherical geometry gives marginally higher optical depths ($\tau=5.4_{-1.4}^{+0.8}$ and 
$\tau=2.6_{-1.2}^{+1.4}$, respectively) for consistent electron temperatures. 

%%%%%%%%%%%%%%%%%%%%%%%%%%%%%%%%%%%%%%%%%%%%%%%%%%%%%%%%%%%%%%%%%%%%
\begin{figure*}[ht!]
\begin{center}
\includegraphics[width=\columnwidth]{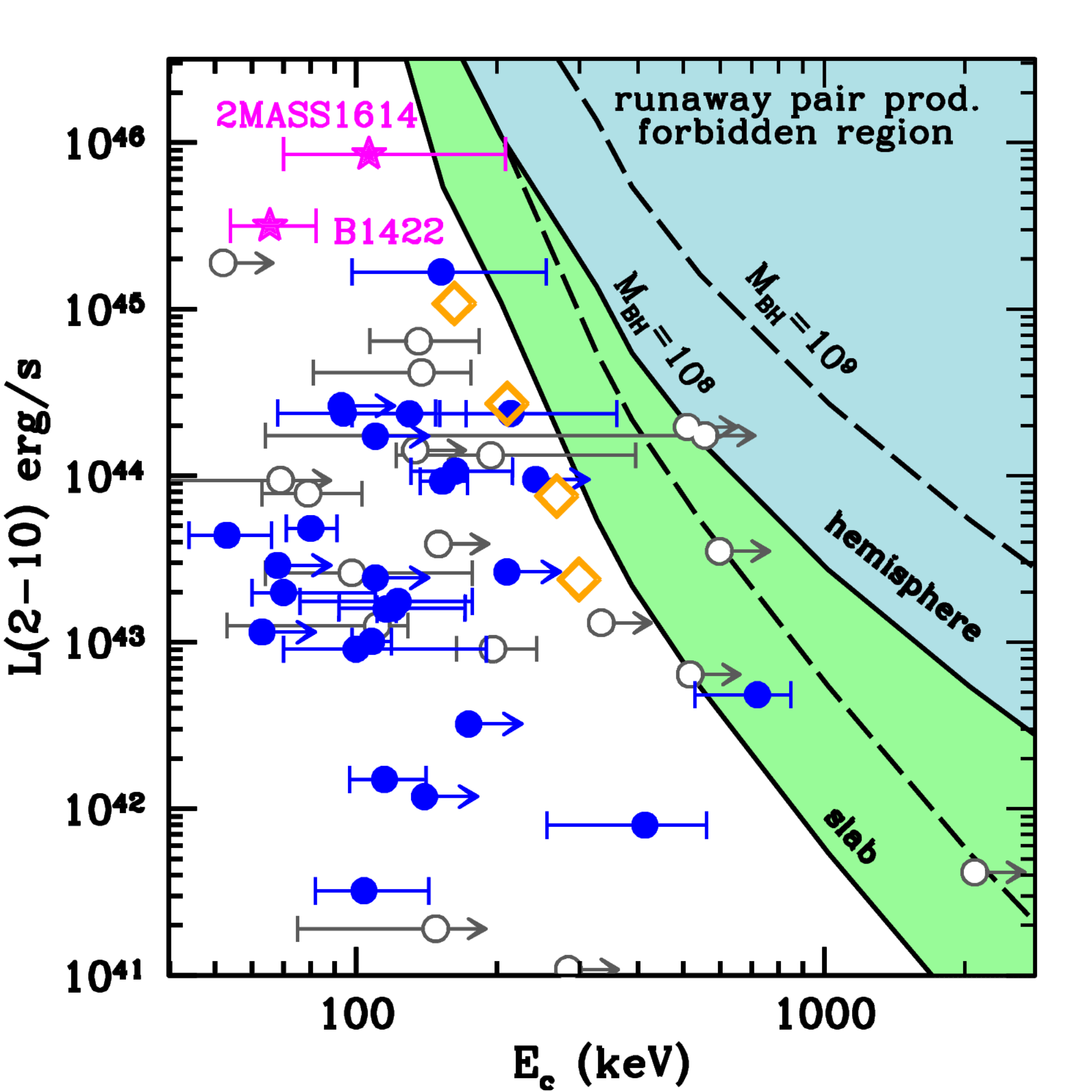}\vspace{0.1cm}
\includegraphics[width=\columnwidth]{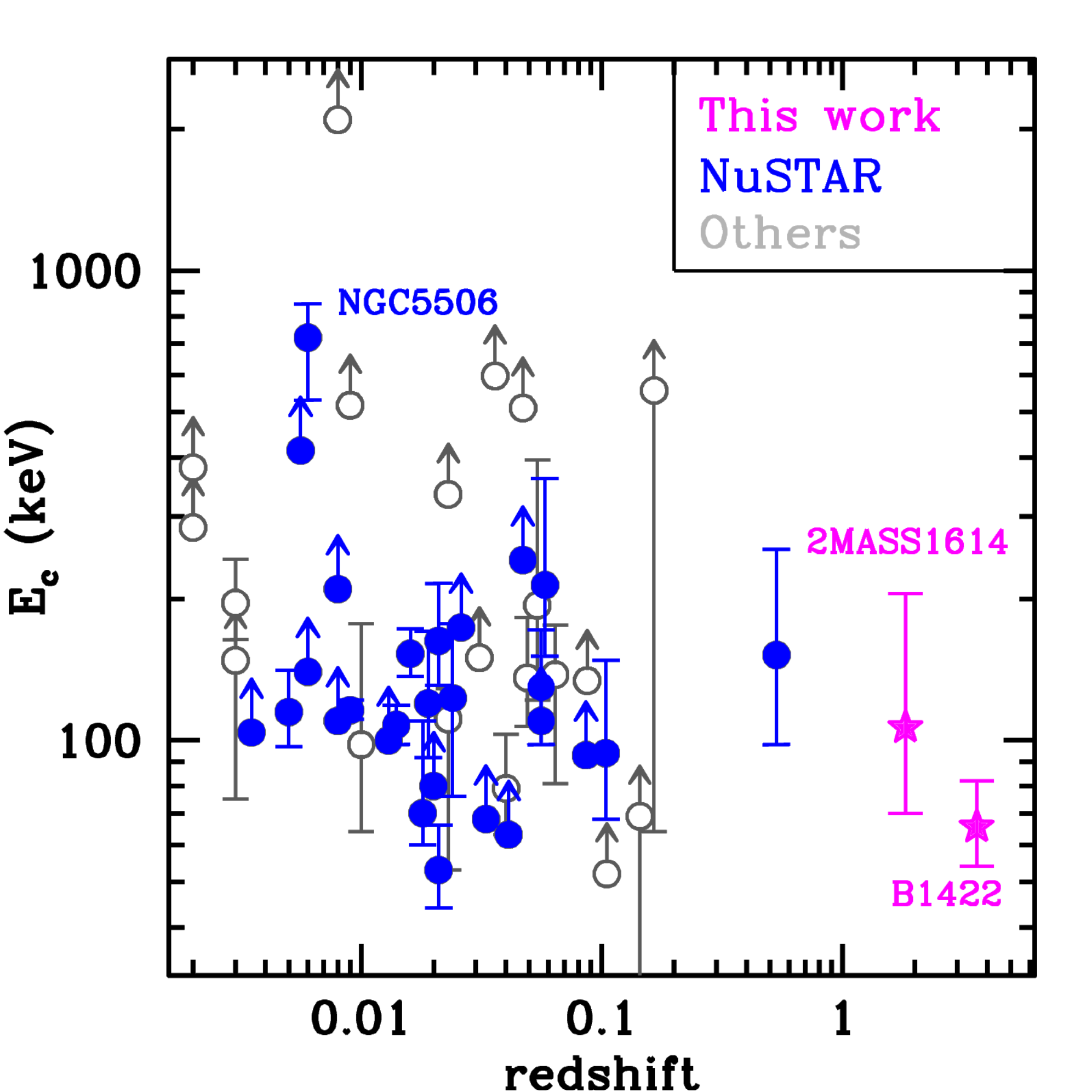}
\caption{{\it Left:}
\ecut\ vs. \lum\ for the updated F15 sample (\nus\ measurements in blue, non-focusing hard X-ray telescopes in gray). The two QSOs analyzed here are in magenta. The cyan (green) area shows the forbidden region due to runaway pair-production for an hemispheric (slab) corona for \mbh$=10^8$ \msun. Dashed lines show the same limits for \mbh$=10^9$ \msun. Orange diamonds show the results from the BAT AGN survey (Ricci et al. 2018). {\it Right:} \ecut\ vs. redshift for the same samples reported in the left panel.}
\label{fig:ctf}
\end{center}
\end{figure*}
%%%%%%%%%%%%%%%%%%%%%%%%%%%%%%%%%%%%%%%%%%%%%%%%%%%%%%%%%%%%%%%%%%%%%%%

Both physical models agree in finding low coronal temperatures, around $kT_e=25-45$ keV, and optical depth substantially above unity, $\tau=1.5-3.5$.
This implies that the appropriate conversion factor between the observed high energy cut-off and the electron temperature should be \ecut$/kT_e\sim3$ instead of 2 for these sources (Petrucci et al. 2001).

\subsection{Absorption model}

For B1422 two possible solutions were explored in Dadina et al. (2016): the reflection and the complex-absorption scenarios. They were both consistent with the \xmm\ data available at that time, from a statistical point of view.
Therefore, we also tested a model in which a complex absorber covers the primary cut-off powerlaw (\texttt{zxipcf$\times$cutoffpl} in Xspec) for both our QSOs. When fitted to the \xmm\ data alone, we obtained in both cases a reasonably good fit with the following parameters: covering factor $f_c\sim0.3-0.4$, absorber column density \nh$\sim(0.5-0.7)\times10^{24}$  \cm2\ and low ionization ($log\xi\sim1.5$ erg cm s$^{-1}$). 
With the addition of \nus\ data, however, it is clear that the reflection model is to be preferred  (see also Risaliti et al. 2013 results on NGC1365), since we measure large $\Delta Cstat$ between the two best-fit models: $\Delta C=24$ for B1422 and $15$ for 2MASSJ16, for two more free parameter.

Interestingly, in the complex-absorption scenario, only lower limits can be derived for the high energy cut-off: \ecut$>235$ keV 2MASSJ16 and \ecut$>171$ keV for B1422, at 90\% c.l. These values, coupled with a higher intrinsic luminosity, by a factor 1.4-1.6, implied by the absorption model, would move both sources within the forbidden region in the $l-T_e$ plane, in tension with the pair-production scenario.

Therefore, an incorrect modeling of the continuum, still consistent with the data below 10 keV for both QSOs, would suggest wrong conclusions about the position of these two QSOs in the pair-production runaway forbidden region, while the addition of \nus\ data clearly rules out this possibility.

\section{Conclusions}

The \ecut\ values measured in this work for two luminous, high-redshift QSOs, fall in the restricted region of the luminosity-temperature plane allowed by runaway pair production at such high luminosities.
Our results on these sources expand by almost one order of magnitude the luminosity range sampled by hard X-ray \ecut\ measurement. The values of \ecut\ derived here are the first ones well determined for luminous QSOs based on \nus\ and \xmm\ quasi-simultaneous data. Consistent results with larger error bars were obtained for B1422 in Dadina et al. (2016) based on \xmm\ data alone, but the lack of hard X-ray data made the results model-dependent (see Sec.~4.3).
A tentative \ecut\ measurement of \ecut$=160_{-80}^{+450}$ keV was presented in Lanzuisi et al. (2016) for PG 1247+267, an hyper-luminous (\lbol$=10^{48}$ erg/s) z=2 QSOs, but the non simultaneity of the soft and hard X-ray data again left the possibility of different interpretations of the combined spectra, giving degenerate results.

Recent results from the Swift-BAT sample (Ricci et al. 2018) show that the average \ecut\ of the sample anti-correlates with the Eddington ratio. At the \edd\ ratio levels of B1422 and 2MASSJ16, \edd=0.2 and 0.7, respectively (see Tab.~1), the average values for the Swift-BAT sample are
\ecut$\sim190$ and $\sim170$ keV. The \ecut\ measured for the two QSOs is therefore lower than for local sources accreting at the same Eddington rate. The main difference is the luminosity range, since $\log (L_{\rm x}/{\rm erg\,s^{-1}})>45$  is not covered by the Swift-BAT sample. 

Our results show that \nus\ has effectively opened a new observational window,
allowing for the first time measurement of coronal temperatures in AGN beyond $z=1$ (Fig.~2, right), where the redshift effect allows, in principle, measurement of \ecut\ values as high as few hundred keV in bright QSOs.
Future observations of other high-redshift, luminous sources will expand the sample of \ecut\ measured in this regime and further test the validity of the pair production model.
To substantially increase the sample over which to test the pair production forbidden region, larger effective area in the hard band (as proposed for HEX-P; see Hickox et al. 2019 WP) is needed to collect enough counts for accurate spectral analysis of faint sources. 

~\\

\acknowledgements
We thank the anonymous referee for her/his very valuable comments.
This work is based on observations obtained
with: the \nus\ mission, a project led by the Caltech, managed by JPL and funded by NASA; \xmm, an ESA science mission funded by ESA and NASA. This research has made use of tools from NASA’s HEASARC,
a service of Goddard SFC and the SAO.  
GL, MC and SB acknowledge financial support from the Italian Space Agency under grant ASI-INAF I/037/12/0, and n. 2017-14-H.O.
FT acknowledges support by the ``Rita Levi Montalcini'' 2014 program.
MG is supported by the Lyman Spitzer Jr. Fellowship (Princeton University) and by NASA Chandra GO7-18121X.

%% This command is needed to show the entire author+affilation list when
%% the collaboration and author truncation commands are used.  It has to
%% go at the end of the manuscript.
%\allauthors

%% Include this line if you are using the \added, \replaced, \deleted
%% commands to see a summary list of all changes at the end of the article.
%\listofchanges


\begin{thebibliography}{}

\bibitem[An et al.(2014)]{2014SPIE.9144E..1QA} An, H., Madsen, K.~K., Westergaard, N.~J., et al.\ 2014, Space Telescopes and Instrumentation 2014: Ultraviolet to Gamma Ray, 91441Q.

\bibitem[Assef et al.(2011)]{2011ApJ...742...93A} Assef, R.~J., Denney, K.~D., Kochanek, C.~S., et al.\ 2011, \apj, 742, 93.

\bibitem[Buisson et al.(2018)]{2018MNRAS.481.4419B} Buisson, D.~J.~K., Fabian, A.~C., \& Lohfink, A.~M.\ 2018, \mnras, 481, 4419.

\bibitem[Cackett et al.(2014)]{2014MNRAS.438.2980C} Cackett, E.~M., Zoghbi, A., Reynolds, C., et al.\ 2014, \mnras, 438, 2980.

\bibitem[Cash(1979)]{1979ApJ...228..939C} Cash, W.\ 1979, \apj, 228, 939.

\bibitem[Cavaliere, \& Morrison(1980)]{1980ApJ...238L..63C} Cavaliere, A., \& Morrison, P.\ 1980, \apj, 238, L63.

\bibitem[Chartas et al.(2009)]{2009ApJ...706..644C} Chartas, G., Saez, C., Brandt, W.~N., et al.\ 2009, \apj, 706, 644.

\bibitem[Chartas et al.(2016)]{2016AN....337..356C} Chartas, G., Rhea, C., Kochanek, C., et al.\ 2016, Astronomische Nachrichten, 337, 356.

\bibitem[Dadina(2008)]{2008A&A...485..417D} Dadina, M.\ 2008, \aap, 485, 417.


\bibitem[Dadina et al.(2016)]{2016A&A...592A.104D} Dadina, M., Vignali, C., Cappi, M., et al.\ 2016, \aap, 592, A104.

\bibitem[De Marco et al.(2013)]{2013MNRAS.431.2441D} De Marco, B., Ponti, G., Cappi, M., et al.\ 2013, \mnras, 431, 2441 

\bibitem[Del Moro et al.(2017)]{2017ApJ...849...57D} Del Moro, A., Alexander, D.~M., Aird, J.~A., et al.\ 2017, \apj, 849, 57.

\bibitem[Fabian et al.(2015)]{2015MNRAS.451.4375F} Fabian, A.~C., Lohfink, A., Kara, E., et al.\ 2015, \mnras, 451, 4375.

\bibitem[Garc{\'\i}a et al.(2014)]{2014ApJ...782...76G} Garc{\'\i}a, J., Dauser, T., Lohfink, A., et al.\ 2014, \apj, 782, 76.

\bibitem[Garc{\'\i}a et al.(2015)]{2015ApJ...813...84G} Garc{\'\i}a, J.~A., Steiner, J.~F., McClintock, J.~E., et al.\ 2015, \apj, 813, 84.

\bibitem[Haardt \& Maraschi(1993)]{1993ApJ...413..507H} Haardt, F. \& Maraschi, L.\ 1993, \apj, 413, 507.

\bibitem[Harrison et al.(2013)]{2013ApJ...770..103H} Harrison, F.~A., Craig, W.~W., Christensen, F.~E., et al.\ 2013, \apj, 770, 103.

%\bibitem[Iwasawa, \& Taniguchi(1993)]{1993ApJ...413L..15I} Iwasawa, K., \& Taniguchi, Y.\ 1993, \apj, 413, L15.

\bibitem[Kammoun et al.(2017)]{2017MNRAS.465.1665K} Kammoun, E.~S., Risaliti, G., Stern, D., et al.\ 2017, \mnras, 465, 1665.

\bibitem[Kara et al.(2016)]{2016MNRAS.462..511K} Kara, E., Alston, W.~N., Fabian, A.~C., et al.\ 2016, \mnras, 462, 511.

\bibitem[Lanzuisi et al.(2016)]{2016A&A...590A..77L} Lanzuisi, G., Perna, M., Comastri, A., et al.\ 2016, \aap, 590, A77.

\bibitem[MacLeod et al.(2015)]{2015ApJ...806..258M} MacLeod, C.~L., Morgan, C.~W., Mosquera, A., et al.\ 2015, \apj, 806, 258.

\bibitem[Magdziarz, \& Zdziarski(1995)]{1995MNRAS.273..837M} Magdziarz, P., \& Zdziarski, A.~A.\ 1995, \mnras, 273, 837.

\bibitem[Malizia et al.(2014)]{2014ApJ...782L..25M} Malizia, A., Molina, M., Bassani, L., et al.\ 2014, \apj, 782, L25.

\bibitem[Matt et al.(2015)]{2015MNRAS.447.3029M} Matt, G., Balokovi{\'c}, M., Marinucci, A., et al.\ 2015, \mnras, 447, 3029.

\bibitem[Molina et al.(2019)]{2019MNRAS.484.2735M} Molina, M., Malizia, A., Bassani, L., et al.\ 2019, \mnras, 484, 2735.

\bibitem[Nandra et al.(2007)]{2007MNRAS.382..194N} Nandra, K., O'Neill, P.~M., George, I.~M., et al.\ 2007, \mnras, 382, 194.

\bibitem[Orienti et al.(2007)]{2007A&A...475..813O} Orienti, M., Dallacasa, D., \& Stanghellini, C.\ 2007, \aap, 475, 813.

\bibitem[Parker et al.(2015)]{2015ApJ...808....9P} Parker, M.~L., Tomsick, J.~A., Miller, J.~M., et al.\ 2015, \apj, 808, 9.

\bibitem[Perola et al.(2002)]{2002A&A...389..802P} Perola, G.~C., Matt, G., Cappi, M., et al.\ 2002, \aap, 389, 802.

\bibitem[Petrucci et al.(2001)]{2001ApJ...556..716P} Petrucci, P.~O., Haardt, F., Maraschi, L., et al.\ 2001, \apj, 556, 716.

\bibitem[Piconcelli et al.(2005)]{2005A&A...432...15P} Piconcelli, E., Jimenez-Bail{\'o}n, E., Guainazzi, M., et al.\ 2005, \aap, 432, 15.

\bibitem[Pooley et al.(2007)]{2007ApJ...661...19P} Pooley, D., Blackburne, J.~A., Rappaport, S., et al.\ 2007, \apj, 661, 19.

\bibitem[Reis, \& Miller(2013)]{2013ApJ...769L...7R} Reis, R.~C., \& Miller, J.~M.\ 2013, \apj, 769, L7.

\bibitem[Ricci et al.(2018)]{2018MNRAS.480.1819} Ricci, C., Ho, L.~C., Fabian, A.~C., et al.\ 2018, \mnras, 480, 1819.

\bibitem[Risaliti et al.(2013)]{2013Natur.494..449R} Risaliti, G., Harrison, F.~A., Madsen, K.~K., et al.\ 2013, \nat, 494, 449.

\bibitem[Shen et al.(2011)]{2011ApJS..194...45S} Shen, Y., Richards, G.~T., Strauss, M.~A., et al.\ 2011, The Astrophysical Journal Supplement Series, 194, 45.

\bibitem[Stern et al.(1995)]{1995ApJ...449L..13S} Stern, B.~E., Poutanen, J., Svensson, R., et al.\ 1995, \apj, 449, L13.


\bibitem[Svensson(1984)]{1984MNRAS.209..175S} Svensson, R.\ 1984, \mnras, 209, 175.


\bibitem[Tamborra et al.(2018)]{2018MNRAS.475.2045T} Tamborra, F., Papadakis, I., Dov{\v c}iak, M., \& Svoboda, J.\ 2018a, \mnras, 475, 2045 

\bibitem[Tamborra et al.(2018)]{2018A&A...619A.105T} Tamborra, F., Matt, G., Bianchi, S., \& Dov{\v c}iak, M.\ 2018b, \aap, 619, A105 

%\bibitem[Titarchuk(1994)]{1994ApJ...434..570T} Titarchuk, L.\ 1994, \apj, 434, 570.

\bibitem[Tortosa et al.(2017)]{2017MNRAS.466.4193T} Tortosa, A., Marinucci, A., Matt, G., et al.\ 2017, \mnras, 466, 4193.

%\bibitem[Tortosa et al.(2018)]{2018A&A...614A..37T} Tortosa, A., Bianchi, S., Marinucci, A., et al.\ 2018, \aap, 614, A37.

\bibitem[Vasudevan et al.(2013)]{2013ApJ...763..111V} Vasudevan, R.~V., Brandt, W.~N., Mushotzky, R.~F., et al.\ 2013, \apj, 763, 111.

\bibitem[Younes et al.(2019)]{2019ApJ...870...73Y} Younes, G., Ptak, A., Ho, L.~C., et al.\ 2019, \apj, 870, 73.

\bibitem[Zappacosta et al.(2018)]{2018ApJ...854...33Z} Zappacosta, L., Comastri, A., Civano, F., et al.\ 2018, \apj, 854, 33.


\bibitem[Zdziarski et al.(1996)]{1996MNRAS.283..193Z} Zdziarski, A.~A., Johnson, W.~N., \& Magdziarz, P.\ 1996, \mnras, 283, 193.

\bibitem[{\.Z}ycki et al.(1999)]{1999MNRAS.309..561Z} {\.Z}ycki, P.~T., Done, C., \& Smith, D.~A.\ 1999, \mnras, 309, 561 


\end{thebibliography}
\end{document}